\documentclass[twocolumn,secnumarabic,amssymb,nobibnotes,nofootinbib,aps,pre]{revtex4-1}
\usepackage[dvipdfmx]{hyperref}
\usepackage{amssymb,amsmath}
\usepackage{graphicx}
\usepackage{bm}
\usepackage{amsfonts}
\usepackage{comment}
\usepackage{here}
\usepackage{color}

\begin{document}
\title{Particle flows around an intruder}

\author{Satoshi Takada}
\email[e-mail: ]{takada@go.tuat.ac.jp}
\affiliation{Institute of Engineering, Tokyo University of Agriculture and Technology, 
2--24--16, Naka-cho, Koganei, Tokyo 184--8588, Japan}
\author{Hisao Hayakawa}
\affiliation{Yukawa Institute for Theoretical Physics, Kyoto University, 
Kitashirakawa Oiwakecho, Sakyo-ku, Kyoto 606--8502, Japan}
\date{\today}
\begin{abstract}
Particle flows injected as beams and scattered by an intruder are numerically studied.
We find a crossover of the drag force from Epstein's law to Newton's law, depending on the ratio of the speed to the thermal speed.
These laws can be reproduced by a simple analysis of a collision model between the intruder and particle flows.
The crossover from Epstein's law to Stokes' law is also found for the low-speed regime as the time evolution of the drag force caused by beam particles.
We also show the existence of turbulent-like behavior of the particle flows behind the intruder with the aid of the second invariant of the velocity gradient tensor and the relative mean square displacement for the high-speed regime and a large intruder.
\end{abstract}
\maketitle

\section{Introduction}
To know fluid flows around an intruder is a fundamental problem \cite{Lamb, Batchelor}.
When the Reynolds number is low, the drag force acting on a spherical intruder obeys Stokes' law in which the drag is proportional to the relative speed between the intruder and the fluid, the viscosity of the fluid, and the radius of the intruder.
On the other hand, the drag force satisfies Newton's law for the high Reynolds number, in which the drag is proportional to the square of the relative speed and the cross section of the intruder.

To understand particle flows is important in various fields \cite{Hirano06, Adil06, Das10, Armitage, Vallado14, Ramsey, Patterson09, Hutzler12, Cho75, Lee87}.
The perfect fluidity is a key concept to understand quark-gluon matter \cite{Hirano06, Adil06}.
The drag coefficient of quarks through the quark-gluon plasma is evaluated theoretically \cite{Das10}.
The drag force acting on dust in the protoplanetary disks is known to show the crossover from Epstein's law \cite{Epstein24}, in which the drag force is proportional to the cross section and the relative moving speed, to Stokes' law as the average size of dust increases \cite{Armitage}.
This problem is also related to the designs of artificial satellites \cite{Vallado14}.
Atomic and molecular beams \cite{Ramsey, Patterson09, Hutzler12} are widely used for nanotechnologies such as a beam epitaxy on a thin film \cite{Cho75}.
We believe that the drag force still satisfies Stokes' law for molecular flows of low Reynolds number  \cite{Vergeles95, Chen06, Li08, Itami15, Asano18}.
In atomic and molecular flows, Stokes' law can be used only for systems in the zero Knudsen number limit, i.e., if the mean free path of molecules is much smaller than the size of the intruder \cite{Cunningham10, Knudsen11, Sone, Takata93, Taguchi15, Taguchi17}.
The correction to Stokes' drag for rarefied gases in the low Knudsen number is theoretically confirmed by the kinetic theory of rarefied gases \cite{Cunningham10, Knudsen11, Sone, Takata93, Taguchi15, Taguchi17}.
The drag force, however, acting on a slowly moving intruder in a crowd of molecules for the high Knudsen number satisfies Epstein's law \cite{Epstein24, Li03}.

There are various experimental and numerical studies on the drag force acting on an intruder in granular flows \cite{Albert99, Chehata03, Wassgren03, Geng04, Bharadwaj06, Potiguar13, Takehara10, Reddy11, Hilton13, Guillard13, Takada17, Candelier10, Gnann, Wang14}.
Note that thermal fluctuations do not play any roles for granular particles. 
A variety of velocity dependences of the drag force is reported depending on the protocols of experiments and simulations.
Granular jet experiments \cite{Cheng07, Ellowitz13} and simulations \cite{Ellowitz13, Sano12, Muller14} suggest that the granular jet flows can be approximated by a perfect fluid model \cite{Ellowitz13, Muller14}.

It is natural to expect that a molecular flow can create turbulence if the corresponding Reynolds number of the flow is sufficiently high.
To verify this conjecture, the K\'{a}rm\'{a}n vortices behind an intruder have already been observed in molecular dynamics (MD) simulations \cite{Komatsu14, Asano18, Rapaport86}.
Nevertheless, there are no published papers, as long as we know, to reproduce a fully developed turbulent flow by MD.

In this paper, we numerically study the drag force acting on a stationary spherical intruder in particle flows by controlling the ratio of the injected speed of the particles to the thermal speed, which is proportional to the sound speed, in terms of the MD.
We also try to reproduce a turbulent-like behavior of particle flows behind the intruder with the aid of the second invariant of the velocity gradient tensor and the relative mean square displacement of two moving particles if the injected speed is much larger than the thermal speed and the size of the intruder is much larger than the molecule size.

The organization of this paper is as follows:
In the next section, we explain our model and setup of our study.
In Sec.\ \ref{sec:Epstein_Newtonian}, we present the results of our simulation for the drag force acting on the intruder based on not too large systems.
In the first part, we show the crossover from Epstein's drag to Newtonian drag depending on the ratio of the colliding speed to the thermal speed. 
In the second part, we discuss how results depend on the boundary condition on the surface of the intruder. 
In Sec.\ \ref{sec:Stokes_Epstein}, we illustrate the existence of a crossover from Epstein's drag to Stokes' drag as time goes on. 
This section consists of two parts.
In the first subsection, we show the numerical results to exhibit the crossover.
In the second subsection, we explain the mechanism of the crossover.
In Sec.\ \ref{sec:turbulent}, we demonstrate the existence of a turbulent-like flow behind the intruder in which the relative motion of two colliding particles is super-ballistic if the size of the intruder is much larger than the molecule size. 
In the first subsection, we illustrate the conditions to observe turbulent-like flows produced by molecule flows.  
We also discuss the angle distribution of scattered particles by the intruder. 
In Sec.\ \ref{sec:conclusion}, we conclude our results.
In Appendix \ref{sec:c}, we examine whether the sound speed is applicable to characterize the drag.
In Appendix \ref{sec:periodic}, we briefly summarize our numerical results when the intruder is initially put inside the beam particles under the periodic boundary condition in the flow direction.

\begin{figure}[htbp]
	\includegraphics[width=\linewidth]{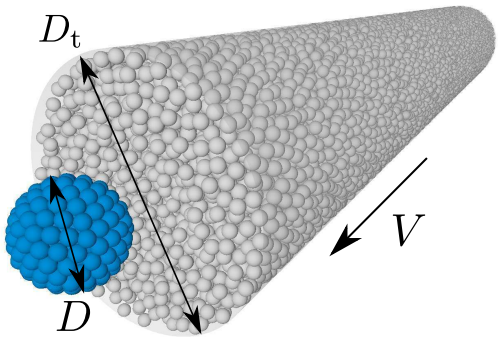}
	\caption{A snapshot of the initial configuration of the system for $\phi=0.40$, $D=7d$, and $D_{\rm t}=20d$.}
	\label{fig:setup}
\end{figure}

\section{Model}\label{sec:model}
In this section, let us explain our model and the setup of our simulation.
The system consists of two parts: one is a fixed intruder, and another is a collection of the mobile particles, as shown in Fig.\ \ref{fig:setup}.
The intruder is made by one core particle whose diameter is $D_{\rm c}$.
In most cases, the intruder is covered by $N_{\rm s}$ identical particles whose diameters are $d$ on the surface of the core particle.
We examine four sizes of the intruders as $(D_{\rm c}/d, N_{\rm s})=(5,144)$, $(13,744)$, $(28, 3{,}364)$, and $(98, 39{,}204)$.
The diameter of the intruder is given by $D=D_{\rm c}+2d$ in this case.
We also examine the case of $N_{\rm s}=0$, i.e., $D=D_{\rm c}$ in Sec.\ \ref{sec:geometry}.
Then, we find that the results are almost independent of the boundary condition on the surface of the intruder. 
The intruder is fixed at the origin, which has an infinitely large mass.
We simulate systems containing $N=30{,}000$, $101{,}250$, $810{,}000$, and $9{,}600{,}000$ monodisperse mobile particles depending on the size of the intruder.
The mass and diameter for each molecule are $m$ and $d$, respectively.
Throughout this paper, collisions between particles are assumed to be elastic.
We examine various initial volume fractions of beam particles ranged from $\phi=0.20$ to $0.55$.
Before starting our simulation, the mobile particles are thermalized with the temperature $T$, and are moved with the translational speed $V$.
The equation of motion of $i$-th particle at the position $\bm{r}_i$ is given by $md^2 \bm{r}_i/dt^2=\sum_j \bm{F}_{ij}$ with the interparticle force $\bm{F}_{ij}=\Theta(d-r_{ij})k (d-r_{ij})\hat{\bm{r}}_{ij}$, where we have introduced $r_{ij}=|\bm{r}_{ij}|$, $\hat{\bm{r}}_{ij}=\bm{r}_{ij}/r_{ij}$, the spring constant $k$, and the step function $\Theta(x)$, i.e., $\Theta(x)=1$ for $x\ge0$ and $\Theta(x)=0$ otherwise.
Initially, we confine the mobile particles in tubes whose diameters are $D_{\rm t}=20d$, $30d$, $60d$, and $200d$ for $D_{\rm c}=5d$, $13d$, $28d$, and $98d$, respectively.
Since the drag force is known to depend on the distance from the boundary of a container through a simulation \cite{Vergeles95} and a theory by fluid mechanics \cite{Brenner61}, we keep the ratio of $D_{\rm t}/D\simeq 2$ in our simulation.
We assume that the interaction force between the wall of the tube and the particles is identical to the interparticle force.
The used parameters are listed in Table \ref{fig:parameters}.

\begin{table}[htbp]
	\centering
	\caption{The set of used parameters.}
 	\begin{tabular}{c|c|c|c}
 	\hline\hline
 	$D_{\rm c}/d$ & $N_{\rm s}$ & $N$ & $D_{\rm t}/d$ \\ \hline
 	$5$ & $144$ & $30{,}000$ & $20$ \\ \hline
 	$13$ & $744$ & $101{,}250$ & $30$ \\ \hline
 	$28$ & $3{,}364$ & $810{,}000$ & $60$ \\ \hline
 	$98$ & $39{,}204$ & $9{,}600{,}000$ & $200$ \\
 	\hline\hline
	\end{tabular}
	\label{fig:parameters}
\end{table}

When the mobile particles collide with the intruder, we examine two cases for the reflection: one is the random reflection of the angle with the temperature $T_{\rm w}$, and another is the simple reflection rule according to the equation of motion. 
For the former condition, when the colliding particle leaves the intruder, we give the velocity to the particle, whose magnitude is the thermal speed $v_{\rm T}=\sqrt{2T/m}$, and the scattering direction of the particle is random on the surface of the intruder.
Here, we set $T_{\rm w}=T$ for simplicity for the former condition.
We also examine two cases for the systems behind the intruder: one is a free scattering case, and another is a confined case, where the scattered particles are still confined in the tube.
In the following, we use the time-averaged drag force in a certain time window, where the instantaneous force acting on the intruder is measured via $\bm{F}=\sum_j\bm{F}_j$, where $\bm{F}_j$ is the contact force acting on the intruder.
In addition, we fix the speed $V=0.10 d\sqrt{k/m}$.
We have verified that the results of our simulation for $V=0.10 d\sqrt{k/m}$ are consistent with those of the hard-core particles.
We also note that the time increment is fixed to be $\Delta t=1.0\times 10^{-3}\sqrt{m/k}$.


\section{Crossover from Epstein's law to Newtonian law}\label{sec:Epstein_Newtonian}
In this section, we present the results of our simulations for not large system sizes.
This section consists of two parts.
In Sec.\ \ref{sec:FV}, we show the drag force against the translational speed as a crossover from Epstein's law to Newtonian law, depending on the ratio of the colliding speed to the thermal speed.
In Sec.\ \ref{sec:geometry}, we investigate whether the geometry of the intruder affects the drag law.

\begin{figure}[htbp]
	\includegraphics[width=\linewidth]{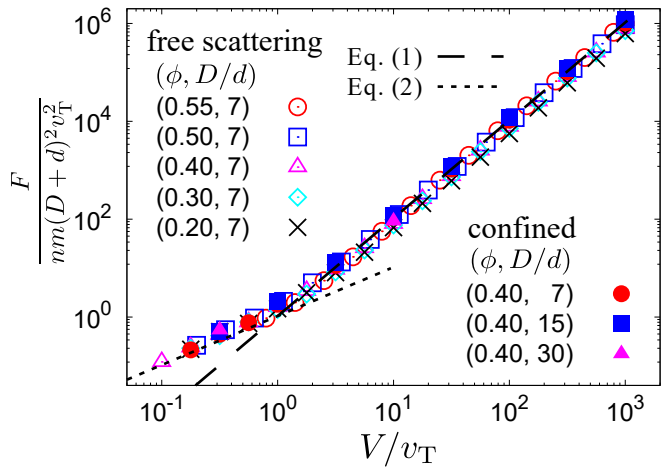}
	\caption{Plots of the drag force against the normalized speed $V/v_{\rm T}$ for the free scattering and the confined cases for various fractions $\phi$ and intruder sizes $D$.
	The dashed and dotted lines represent the collision models \eqref{eq:F_coll} and \eqref{eq:F_coll_low}, respectively.}
	\label{fig:F_vs_V}
\end{figure}

\subsection{Crossover of the drag force from Epstein's law to Newtonian law}\label{sec:FV}
We present the results of our simulation of the drag force acting on the intruder against the normalized colliding speed by the thermal speed for both the free scattering (with fixing $D/d$ and various $\phi$) and the confined cases (with fixing $\phi$ and various $D/d$) in Fig.~\ref{fig:F_vs_V}.
Here, the data are obtained by the time average of the instantaneous drag force in the range between $t=400\sqrt{m/k}$ and $800\sqrt{m/k}$.
Because the drag force is determined from the total force acting on the intruder by all collisions of particles in front of the intruder, the drag force is unchanged whether the particle flow behind the intruder is confined or not.
In this subsection, we assume that the intruder is covered by small particles whose sizes are the same as those of colliding particles and the reflection rule between colliding particles and the intruder is thermal in which reflecting particles are stochastically scattered with satisfying the Maxwell distribution.

\begin{figure}[htbp]
	\includegraphics[width=\linewidth]{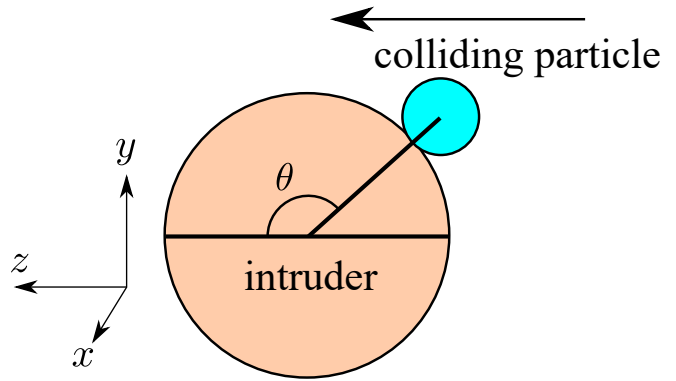}
	\caption{Explanation of the angle $\theta$ between the intruder and the colliding particle.
	The arrow indicates the flow direction.
	We define $z$-axis as the flow direction and the center of the intruder is set as the origin.}
	\label{fig:def_theta}
\end{figure}

We note that our obtained drag forces are proportional to $(D+d)^2$, i.e., the collision cross section \cite{Wassgren03, Takada19}.
The drag force is proportional to the square of the speed for $V/v_{\rm T}\gg 1$, while it is proportional to $v_{\rm T}V$ for $V/v_{\rm T}\ll 1$.

The former, on the one hand, can be understood by a simple collision model.
Because each momentum change in the flow direction for a collision is $\Delta p_\parallel =2mV\cos^2\theta$ where $\theta$ is the angle between the intruder and the colliding particle as shown in Fig.\ \ref{fig:def_theta}, and the collision frequency is $\Omega_{\rm c}=(\pi/4)n(D+d)^2V$ with the number density $n=6\phi/(\pi d^3)$ \cite{Wassgren03}, the force acting on the intruder is given by  
\begin{equation}
	F=\int_0^\pi \sin\theta d\theta \Delta p_\parallel \Omega_{\rm c}
	=\frac{\pi}{3}nm (D+d)^2 V^2,
	\label{eq:F_coll}
\end{equation}
which agrees well with the simulation results for $V\gtrsim v_{\rm T}$ (see Fig.\ \ref{fig:F_vs_V}).
This is the simple Newtonian drag law for the high-speed regime.
\begin{figure}[htbp]
	\includegraphics[width=\linewidth]{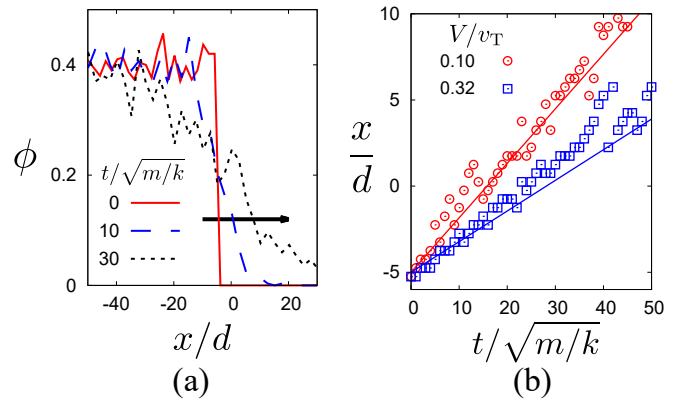}
	\caption{(a) The time evolution of the density profile at $t/\sqrt{m/k}=0$ (red solid line), $10$ (blue dashed line), and $30$ (black dotted line).
	The arrow indicates the direction of the diffusive front.
	(b) The expansion wave ($\phi=0.15$) for $V/v_{\rm T}=0.10$ (open circles) and $0.32$ (open squares).
	The corresponding solid lines represent the expansion speeds $V_{\rm p}$.}
	\label{fig:expansion}
\end{figure}

On the other hand, the front of the beam of mobile particles diffuses before colliding with the intruder for $V/v_{\rm T}\ll 1$ as shown in Fig.\ \ref{fig:expansion}(a).
This expansion speed of the diffusive front is approximately given by $V_{\rm p}=\sqrt{V v_{\rm T}}$ as shown in Fig.\ \ref{fig:expansion}(b).
When we replace $V$ in Eq.\ \eqref{eq:F_coll} by $V_{\rm p}$, we obtain
\begin{equation}
	F=\frac{\pi}{3}nm (D+d)^2 v_{\rm T}V,
	\label{eq:F_coll_low}
\end{equation}
which agrees well with the simulation results for $V\lesssim v_{\rm T}$ as shown in Fig.\ \ref{fig:F_vs_V}.
Here, the sound speed is another characteristic speed that appears in this system.
However, it is evident that the thermal speed is superior to the sound speed to characterize the drag force as shown in Appendix \ref{sec:c}.

\begin{figure}[htbp]
	\includegraphics[width=\linewidth]{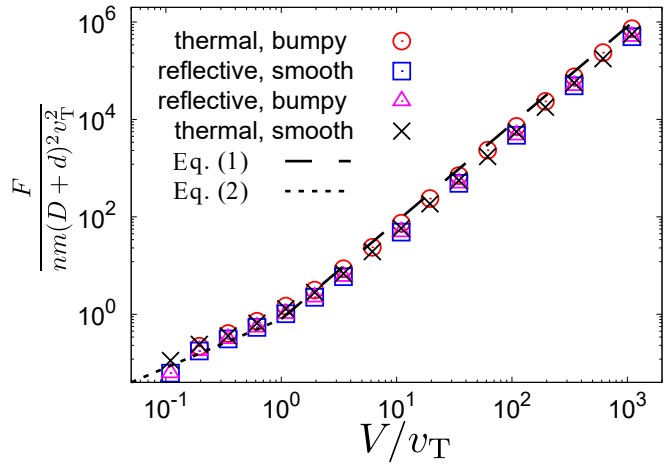}
	\caption{Plots of the drag force against the scaled collision speed $V/v_{\rm T}$ for the free scattering case with $\phi=0.40$ and $D=7d$ when we examine several reflection rules between the intruder and the mobile particles.}
	\label{fig:F_vs_V_boundary}
\end{figure}

\subsection{Effects of the boundary conditions on the intruder}\label{sec:geometry}
In this subsection, let us check whether the results depend on the boundary conditions between the intruder and the mobile particles.
Here, we refer to the intruder introduced in the previous subsection as (i) ``thermal and bumpy'', because the mobile particles reflect at random when they collide with the intruder, and the small particles are attached on the surface of the core particle.
We examine the other three different types of intruders: (ii) ``reflective and smooth,'' (iii) ``reflective and bumpy,'' and (iv) ``thermal and smooth.''
The intruder in the condition of (ii) ``reflective and smooth'' consists of only one core particle, and the reflection between the intruder and the mobile particles is described by a simple elastic collision rule.
The intruder in (iii) ``reflective and bumpy'' adopts a simple elastic collision rule, and the intruder consists of the core particle and the small particles on its surface as that used in the case (i).
The intruder in (iv) ``thermal and smooth'' adopts the thermal scattering as used in (i), and the intruder consists of only one core intruder.
Figure \ref{fig:F_vs_V_boundary} plots the results of the drag forces under various boundary conditions for $D/d=7$ and $\phi=0.40$ in the free scattering condition.
The results clearly indicate that the choice of the boundary condition on the intruder is not important for the drag force.

\section{Crossover from Epstein's law to Stokes' law}\label{sec:Stokes_Epstein}
In this section, we present the crossover from Epstein's law to Stokes' law for the low-speed regime.
This section consists of two parts. 
In the first subsection, we present numerical results to exhibit the crossover from Epstein's law to Stokes' law.
In the second subsection, we explain the physical mechanism of this crossover.  

\begin{figure*}[htbp]
	\includegraphics[width=\linewidth]{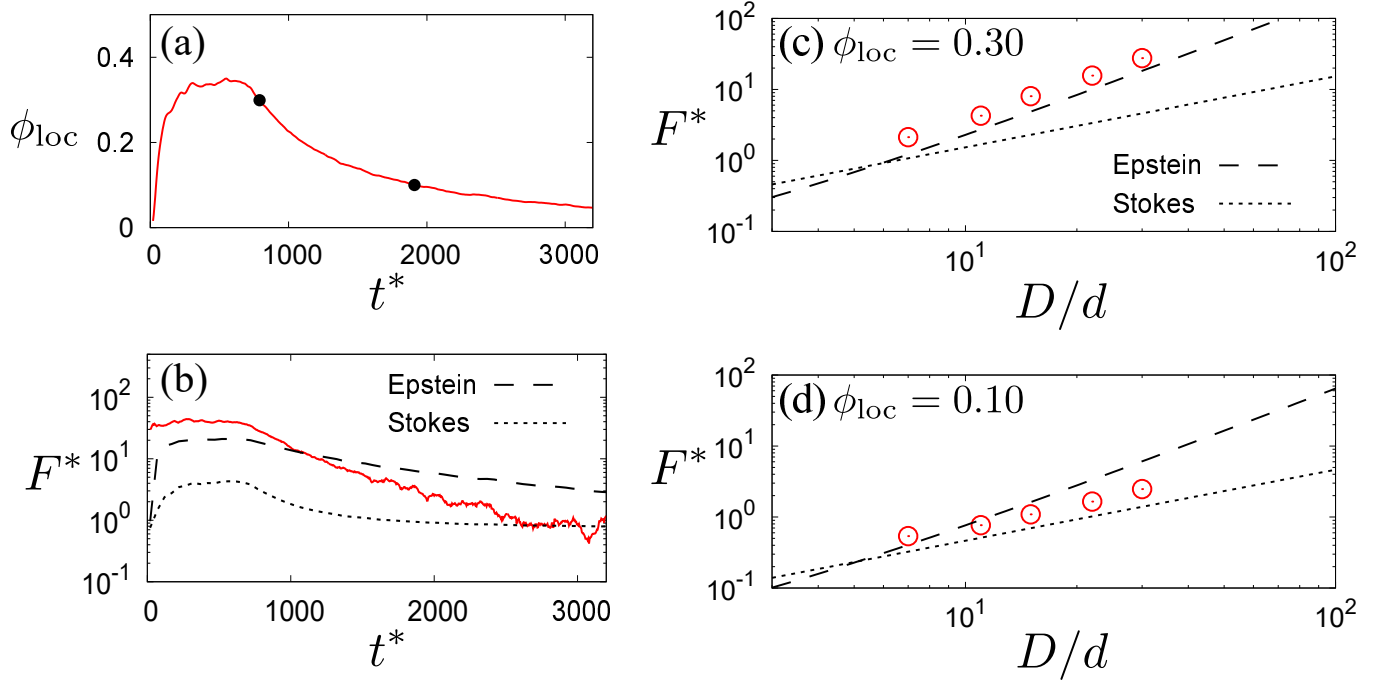}
	\caption{(a) Time evolution of the local packing fraction $\phi_{\rm loc}$ in front of the intruder for $\phi=0.40$, $D/d=30$, and $V/v_{\rm T}=0.32$.
	Two black dots represent the points for $\phi_{\rm loc}=0.30$ and $0.10$, respectively.
	(b) Time evolution of the force acting on the intruder.
	We also plot Epstein's law \eqref{eq:F_coll_low} (dashed line) and Stokes' law \eqref{eq:Stokes} (dotted line) using the time evolution of the local packing fraction $\phi_{\rm loc}$.
	The diameter dependence of the force for (c) $\phi_{\rm loc}=0.30$ and (d) $0.10$, respectively.
	Here, we have introduced the dimensionless force $F^*\equiv F/(kd)$ and time $t^*\equiv t/\sqrt{m/k}$, respectively.}
	\label{fig:Epstein_Stokes}
\end{figure*}

\subsection{How can we get crossover from Epstein's law to Stokes' law?}
In the previous section, we have reported the crossover from Epstein's law to Newtonian law when we control the ratio of the colliding speed to the thermal speed $V/v_{\rm T}$.
Although Newtonian law for $V/v_{\rm T}\gg 1$ is expected,\footnote{We have confirmed that the drag law in this regime remains Newtonian in the range of $200\sqrt{m/k}\lesssim t\lesssim 1500\sqrt{m/k}$.} Epstein's law for $V/v_{\rm T}\ll 1$ is a little unexpected because Stokes' law might be expected for slow flows. 
To clarify the condition to emerge Stokes' law, in this section, we illustrate the existence of a crossover from Epstein's law to Stokes' law with the time evolution by fixing $V/v_{\rm T}\ll 1$.

When we focus on the region in front of the intruder, the packing fraction of this local region changes as time goes on (see Fig.\ \ref{fig:Epstein_Stokes}(a)).
Here, we measure the local packing fraction in the region $-D/2-10d\le z\le -D/2$ (see Fig.\ \ref{fig:def_theta} for the information of the geometry).
After the early stage of the collision process, there is a relatively long metastable state in which the local packing fraction is almost equal to the initial packing fraction of the beam particles. 
Then, $\phi_{\rm loc}$ decreases with time for $t/\sqrt{m/k}\gtrsim 600$.
When we substitute the local packing fraction $\phi_{\rm loc}(t)$ into Eq.\ \eqref{eq:F_coll_low}, we can observe the crossover from Epstein's drag to Stokes' drag as shown in Fig.\ \ref{fig:Epstein_Stokes}(b).
Then, the drag force reaches Stokes' law in the late stage, which is given by
\begin{equation}
	F_{\rm St}=3\pi \eta D V,
	\label{eq:Stokes}
\end{equation}
where the shear viscosity $\eta$ is estimated from the well-known result from the Enskog theory \cite{Ferziger, Garzo99} as
\begin{align}
	\eta(\phi)
	&=\frac{5}{16d^2}\sqrt{\frac{mT}{\pi}}\nonumber\\
	&\hspace{1em}\times \left[\frac{1}{g_0(\phi)}\left(1+\frac{8}{5}\phi g_0(\phi)\right)^2 + \frac{768}{25\pi}\phi^2g_0(\phi)\right].
\end{align}
Here, $g_0(\phi)$ is the radial distribution function at contact, which is approximately given by Carnahan and Starling formula $g_0(\phi)=(1-\phi/2)/(1-\phi)^3$, which is valid for $\phi<0.49$ \cite{Carnahan69}.
We also confirm that the drag force is proportional to the cross section in Epstein's regime, while it is proportional to $D$ in Stokes' regime as shown in Figs.\ \ref{fig:Epstein_Stokes}(c) and (d).
This is also another evidence of the crossover from Epstein's law to Stokes' law.
We note that Stokes' law is also observed when we adopt the periodic boundary condition in the flow direction, which is investigated in Appendix \ref{sec:periodic}.

\subsection{Mechanism of the crossover}
Let us discuss the reason why the time-dependent crossover from Epstein's to Stokes' laws appears.
The flow is stacked around the intruder because the particles cannot move due to the existence of the outer particles near the intruder.
These stacked particles can be regarded as a ``boundary layer'' around the intruder, which prohibits mobile particles from direct collisions with the intruder.
In this case, the local shear rate can be approximated as $\dot\gamma_{\rm loc}\simeq (V/D)\sin\theta$, where $\sin\theta$ represents the projection parallel to the tangential direction of the surface.
This leads to that the shear stress acts on the intruder whose magnitude in the flow direction is $\dot\gamma_{\rm loc}\eta \sin\theta$ if the viscosity $\eta$ is well defined.
Therefore, we can evaluate the force acting on the intruder by integrating the shear stress over the surface of the intruder, and we obtain
\begin{equation}
	F= 2\pi \eta \left(\frac{D}{2}\right)^2 \int_0^\pi d\theta \dot\gamma_{\rm loc}\sin\theta 
	\sim \eta D V,
\end{equation}
which is the origin of Stokes' law.

The particles, however, are not stacked near the intruder in the early stage of the simulation.
Therefore, the beam particles can directly collide with the intruder, whose speed is approximately given by the thermal speed.
Then, we obtain Epstein's law as in Eq.\ \eqref{eq:F_coll_low}.

\section{Turbulent-like flows for $V/v_{\rm T}\gg 1$ and $D/d\gg 1$}\label{sec:turbulent}

In Sec. III, we did not discuss what flows can be observed for $V/v_{\rm T} \gg 1$.
It is natural to expect turbulent-like flows can be observed in this regime, because $V/v_{\rm T}$ might correspond to the Reynolds number of the fluid flows.
We also examine the effect of $D/d$ which has not been investigated in the previous sections.

\begin{figure}[htbp]
	\includegraphics[width=\linewidth]{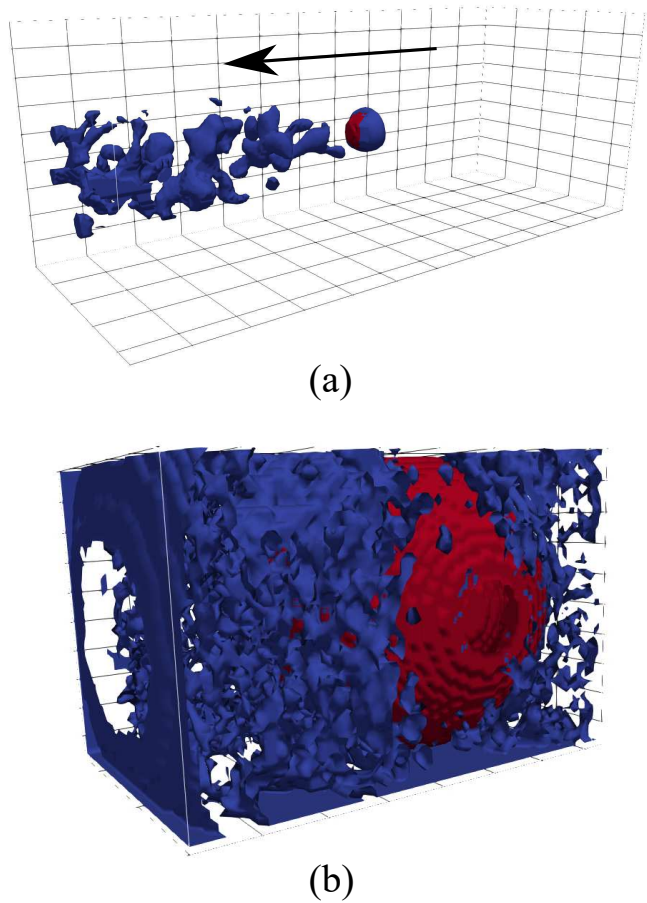}
	\caption{(a) A contour plot of the second invariant $Q=0$ for $\phi=0.40$, $D=7d$, and $V/v_{\rm T}=10$.
	The arrow indicates the flow direction.
	(b) The corresponding plot for $\phi=0.40$, $D=100d$, and $V/v_{\rm T}=10^3$.}
	\label{fig:Q}
\end{figure}

Let us characterize the particle flows behind the intruder for large $V/v_{\rm T}$.
We introduce the second invariant of the velocity gradient tensor $Q=(1/2)(-S_{\alpha\beta}S_{\alpha\beta}+W_{\alpha\beta}W_{\alpha\beta})$ where $S_{\alpha\beta}=(1/2)(\partial_\beta u_\alpha + \partial_\alpha u_\beta)$ and $W_{\alpha\beta}=(1/2)(\partial_\beta u_\alpha - \partial_\alpha u_\beta)$ \cite{Hunt88}.
Here, we adopt Einstein's rule for $\alpha$ and $\beta$ where duplicated indices take summation over $x$, $y$, and $z$.
Figure \ref{fig:Q} shows the contours of $Q=0$ for (a) $D=7d$ and (b) $D=100d$, where the field is coarse-grained with the scale $2d$ for visibility \cite{Zhang10}. 
The vortex rich regions $Q>0$ emerge behind the intruder.
Such domain structure becomes complicated for large $D/d$.
This behavior is similar to that observed in a turbulent flow induced by an intruder \cite{Ott00}.
In particular, huge number of vortex rich domains appear for $D/d\gg 1$ (Fig.\ \ref{fig:Q}).
This suggests that the particles flow in Fig.\ \ref{fig:Q}(b) is turbulent-like.

\begin{figure*}[htbp]
	\includegraphics[width=0.8\linewidth]{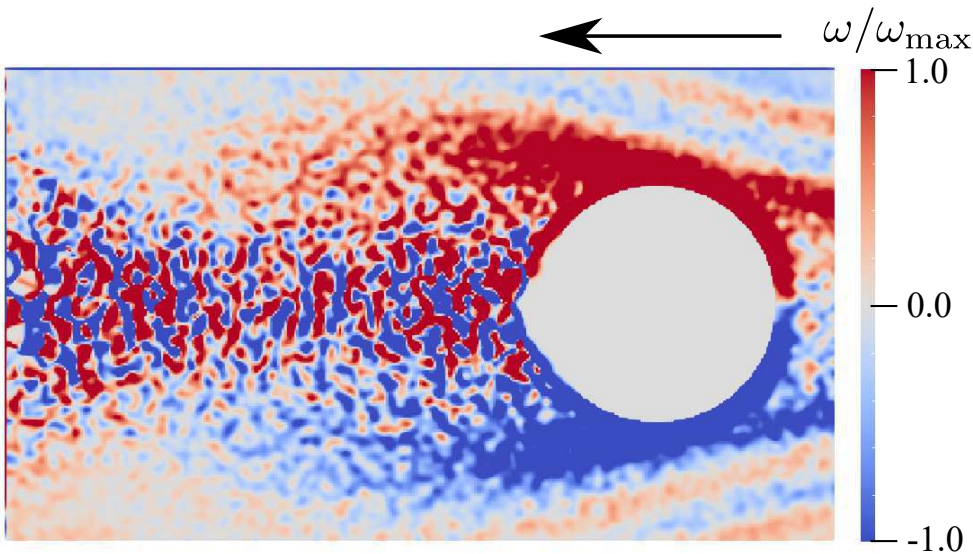}
	\caption{Vorticity field behind the intruder for $V/v_{\rm T}=1.0\times10^3$ and $D/d=100$.
	The color represents the value of $\omega_{xz}/\omega_{\rm max}$ with $\omega_{\rm max}=1.0\times 10^{-3}\ (k/m)^{1/2}$.
	The arrow indicates the flow direction.}
	\label{fig:vorticity}
\end{figure*}

We also characterize the vortex structure by plotting the vorticity induced by the scattering of the intruder.
Because the twisted structure of the flow field is not observed in our simulation, we focus on the flow structure in $xz$-plane (see Fig.\ \ref{fig:def_theta} for the information of the geometry).
Let us introduce the vorticity in $xz$-plane with $|y_i|\le D/2$ as
\begin{equation}
	\omega_{xz}=\frac{\partial v_z}{\partial x} - \frac{\partial v_x}{\partial z}.
\end{equation}
Figure \ref{fig:vorticity} shows a typical snapshot of the vorticity field in $xz$-plane for $D=100d$.
The positive and negative vorticity regimes are generated in the vicinity of the intruder and move toward the downstream.
See also the movie in the Supplemental Material \cite{Movie}.
These complicated structures of vorticities are also similar to those observed in turbulent flows.

\begin{figure}[htbp]
	\includegraphics[width=\linewidth]{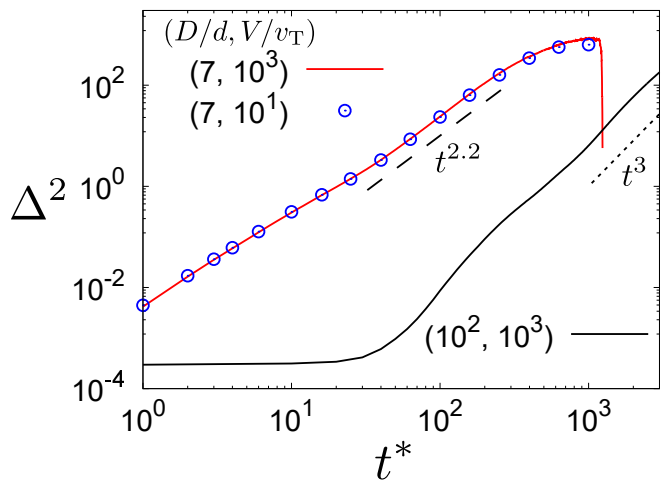}
	\caption{The mean square displacement between two particles for $D/d=7$ with $V/v_{\rm T}=1.0\times 10^3$ and $10.0$, and $D/d=100$ and $V/v_{\rm T}=1.0\times 10^3$.
	The dashed and dotted lines are the guide lines for the exponent $\beta=2.2$ and $3$, respectively.}
	\label{fig:relative_diffusion}
\end{figure}

Let us study how the particles are scattered after collisions by the intruder.
We focus on the relative motion of the particles which collide with the intruder almost simultaneously through the mean square displacement
\begin{equation}
	\Delta^2(t) 
	\equiv \langle |\delta \bm{r}_i(t)-\delta \bm{r}_j(t)|^2 \rangle / d^2,
	\label{eq:Delta}
\end{equation}
with $\delta \bm{r}_i(t)\equiv \bm{r}_i(t+t_{\rm c}) -\bm{r}_i(t_{\rm c})$, where we only select two particles ($i$ and $j$) within the interval $|t_i-t_j|<\Delta t_{\rm th}\equiv 10 \sqrt{m/k}$ with the collision times $t_i$ and $t_j$ with the intruder for $i$-th and $j$-th particles, respectively.
Note that we choose $t_{\rm c}$ in $\delta \bm{r}_i(t)$ or $\delta \bm{r}_j(t)$ in Eq.\ \eqref{eq:Delta} to be much larger time of $t_i$ and $t_j$.
This means that these two particles exist sufficiently far from the intruder at initial.
We also note that we select particles whose position in the $z$-direction is located in $-135\le z_i/d\le-100$ at time $t=0$ for $D/d=100$.
This is because we try to clarify how localized particles behave as time goes on.
Figure \ref{fig:relative_diffusion} shows the super-ballistic behavior $\Delta^2(t)\sim t^{\beta}$ where the best fitted values of exponent $\beta$ are $\beta=2.2$ for $D=7d$ and $\beta=3.0$ for $D=100d$, respectively.
We have checked that this result is insensitive to the choice of $\Delta t_{\rm th}$ in the range $1.0 \sqrt{m/k}\lesssim \Delta t_{\rm th}\lesssim 20\sqrt{m/k}$.
The behavior for $D=100d$ is analogous to the relative motion of two tracer particles in turbulent flows, which is known as Richardson's law $\Delta^2(t)\sim t^3$ \cite{Richardson26, Monin, Beffetta02}.
This suggests that the flow is in a fully turbulent state for $D/d=100$.

\begin{figure}[htbp]
	\includegraphics[width=\linewidth]{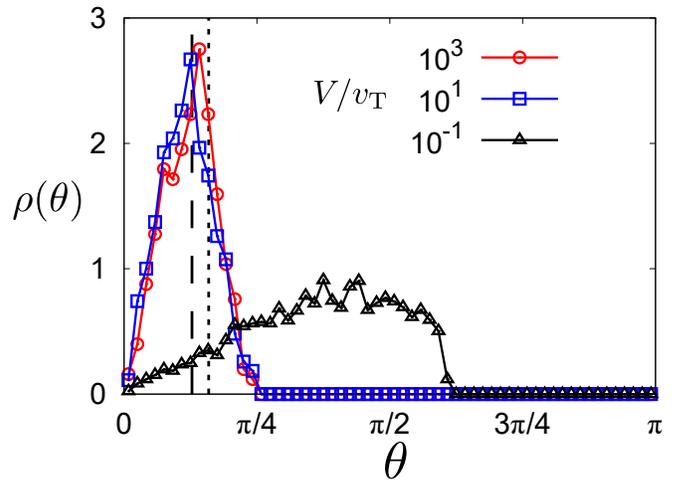}
	\caption{The angular distribution function of the scattered particles for (a) $V/v_{\rm T}=1.0\times 10^3$ (open circles), (b) $10$ (open squares), and (c) $0.10$ (open triangles).
	The dashed and dotted lines represent the opening angles obtained from the simulation and Eq.\ \eqref{eq:theta0}, respectively.
	Here, the definition of the angle $\theta$ is equivalent to that in Fig.\ \ref{fig:def_theta}.}
	\label{fig:scattering_angle}
\end{figure}

Next, we consider the scattering angle distribution of the mobile particles for $D=7d$.
The azimuthal angle is stored when the particles reach the region $|\bm{r}_i|=15d$.
The behavior of the angular distribution $\rho(\theta)$ for $V/v_{\rm T}\gtrsim 1$ completely differs from that for $V/v_{\rm T}\lesssim 1$ as shown in Fig.\ \ref{fig:scattering_angle}.

\begin{figure}[htbp]
	\includegraphics[width=\linewidth]{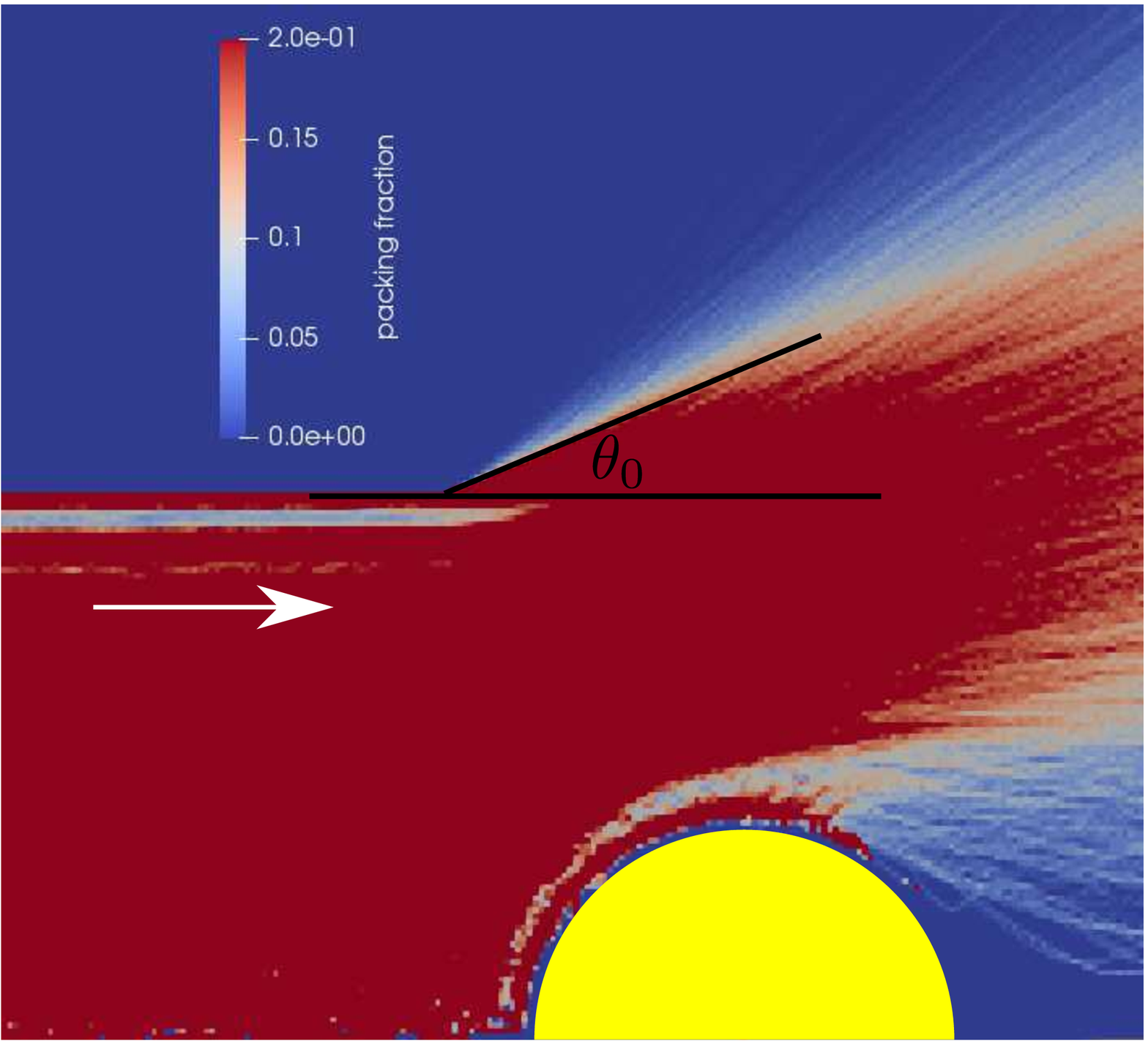}
	\caption{The definition of the open angle $\theta_0$.
	We choose the opening angle as the direction where the volume fraction is $0.15$.
	The large particle represents the intruder.
	The arrow indicates the flow direction.}
	\label{fig:def_angle}
\end{figure}

For $V/v_{\rm T}\gg 1$, the angular distribution of scattered particles has a sharp peak around the opening angle, which is the half of the apex angle of the cone of the beam scattered after collisions with the intruder \cite{Cheng07, Sano12} as shown in Fig.\ \ref{fig:def_angle}.
We also note that this opening angle can be explained by phenomenology as in Ref.\ \cite{Cheng07}.
Because we use repulsive and elastic particles, the opening angle is expressed as 
\begin{equation}
	\theta_0 = \cos^{-1} \left[1-\left(\frac{D}{D_{\rm t}}\right)^2\right],
	\label{eq:theta0}
\end{equation}
for $D<D_{\rm t}$, where $D_{\rm t}$ is the radius of tube of beam particles (see Fig.\ \ref{fig:setup}) \cite{Cheng07}.
When we substitute $D=7d$ and $D_{\rm t}=20d$ into Eq.\ \eqref{eq:theta0}, we obtain $\theta_0=0.50$ (rad), which roughly agrees with the simulation results (see Fig.\ \ref{fig:scattering_angle}).
For $V/v_{\rm T}\ll 1$, on the other hand, the particles are scattered in various directions, as shown in Fig.\ \ref{fig:scattering_angle}.

\section{Conclusion}\label{sec:conclusion}

We have numerically studied the particle flows injected as a beam and scattered by a spherical intruder.
We have found the crossover from Epstein's law to Newton's law, depending on the ratio of the speed to the thermal speed $V/v_{\rm T}$.
This crossover can be explained by a simple collision model.
The crossover from Epstein's law to Stokes' law is also verified as the time evolution of the drag force acting on the intruder.
The turbulent-like behavior has also been observed for $V/v_{\rm T}\gg 1$ and $D/d\gg 1$, where the relative displacement between two tracer particles is super-ballistic, which satisfies Richardson's law.

Although we mainly stress that $V/v_{\rm T}$ is an important control parameter to characterize the particles flows in this paper, it is obvious that $D/d$ is also another important parameter to characterize the flows, particularly for turbulent-like flows. 
The systematic studies for $D/d$ dependent flows will be one of our future research.

\section*{Acknowledgments}

The authors thank Hiroshi Watanabe, Kuniyasu Saitoh, Takeshi Kawasaki, and Michio Otsuki for their useful comments.
The work of S.T.\ has been partially supported by the Grant-in-Aid of MEXT for Scientific Research (Grant No.\ 20K14428).
The research of H.H.\ has been partially supported by the Grant-in-Aid of MEXT for Scientific Research (Grant No.\ 16H04025) and the Scholarship ISHIZUE 2020 of Kyoto University Research Development Program.
The research of the authors was partially supported by the YITP activities (YITP--T--18--03 and YITP--W--18--17).
Numerical computation in this work was partially carried out at the Yukawa Institute Computer Facility. 

\appendix
\begin{figure}[htbp]
	\includegraphics[width=\linewidth]{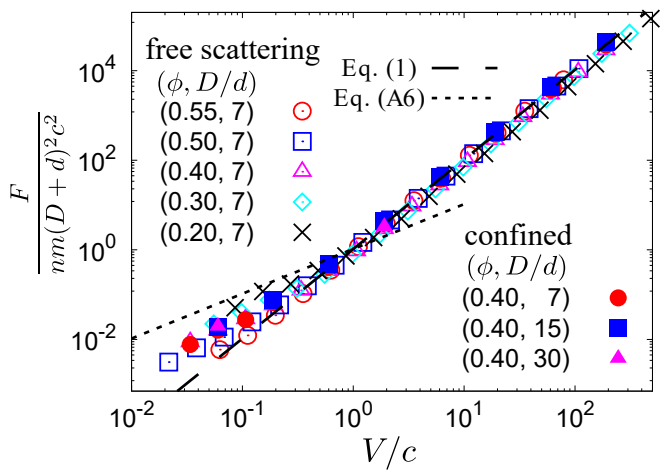}
	\caption{Plots of the drag force against the dimensionless velocity characterized by the sound speed \eqref{eq:c} for free scattering and confined cases for various $\phi$ and intruder sizes $D$.
	The dashed and dotted lines represent the collision model \eqref{eq:F_coll} and \eqref{eq:F_cV}, respectively.}
	\label{fig:FV_c}
\end{figure}
\section{Volume fraction dependence of the sound speed}\label{sec:c}

In this Appendix, we show how the sound speed depends on the volume fraction.
We also examine whether $v_{\rm T}$ can be replaced by the sound speed.

Because the volume fraction of the mobile particles is finite, the equation of state deviates from that for the ideal gas.
Thus, the equation of state at finite density is fitted by the Carnahan--Starling equation \cite{Carnahan69}
\begin{equation}
	\frac{p}{nT}=Z(\phi)=\frac{1+\phi+\phi^2-\phi^3}{(1-\phi)^3},
	\label{eq:CS}
\end{equation}
where $p$ is the pressure.
For the adiabatic process, the first law of thermodynamics becomes
\begin{equation}
	C_V dT + T\left(\frac{\partial p}{\partial T}\right)_V \frac{dL^3}{N}=0,
	\label{eq:1st_law}
\end{equation}
where $L^3$ is the volume, $C_V=3/2$ is the heat capacity at constant volume.
Substituting the equation of state \eqref{eq:CS} into Eq.\ \eqref{eq:1st_law}, the following quantity is conserved along the streamline:
\begin{equation}
	\log \frac{p}{\phi Z(\phi)} - \frac{2}{3}\int^\phi \frac{Z(\phi^\prime)}{\phi^\prime}d\phi^\prime = {\rm const.}
\end{equation}
Then, the sound speed in the adiabatic process is given by
\begin{equation}
	c(\phi)=\sqrt{\left(\frac{\partial p}{\partial \rho}\right)_S}=f(\phi)\sqrt{\frac{T}{m}},
	\label{eq:c}
\end{equation}
with
\begin{equation}
	f(\phi)=\sqrt{ \frac{5+10\phi-3\phi^2-24\phi^3+37\phi^4-22\phi^5+5\phi^6}{3(1-\phi)^6}}.
\end{equation}
If we adopt the expansion speed $\sqrt{c(\phi)V}$, the drag force is given by
\begin{equation}
	F=\frac{\pi}{3}nm(D+d)^2 cV,
	\label{eq:F_cV}
\end{equation}
which cannot capture the simulation results for the low $V/c$ as shown in Fig.\ \ref{fig:FV_c}, where $c/v_{\rm T}=2.07$ ($\phi=0.20$), $3.22$ ($\phi=0.30$), $5.25$ ($\phi=0.40$), $9.21$ ($\phi=0.50$), and $12.7$ ($\phi=0.55$), respectively.

\begin{table}[htbp]
	\centering
	\caption{The set of used parameters in Appendix \ref{sec:periodic}.}
 	\begin{tabular}{c|c|c|c}
 	\hline\hline
 	$D_{\rm c}/d$ & $N_{\rm s}$ & $N$ & $D_{\rm t}/d$ \\ \hline
 	$5$ & $144$ & $30{,}000$ & $20$ \\ \hline
 	$9$ & $400$ & $30{,}000$ & $40$ \\ \hline
 	$13$ & $784$ & $30{,}000$ & $40$ \\ \hline
 	$16$ & $1{,}156$ & $30{,}000$ & $40$ \\ \hline
 	$19$ & $1{,}600$ & $30{,}000$ & $40$ \\
 	\hline\hline
	\end{tabular}
	\label{fig:parameters2}
\end{table}
\begin{figure}[htbp]
	\includegraphics[width=\linewidth]{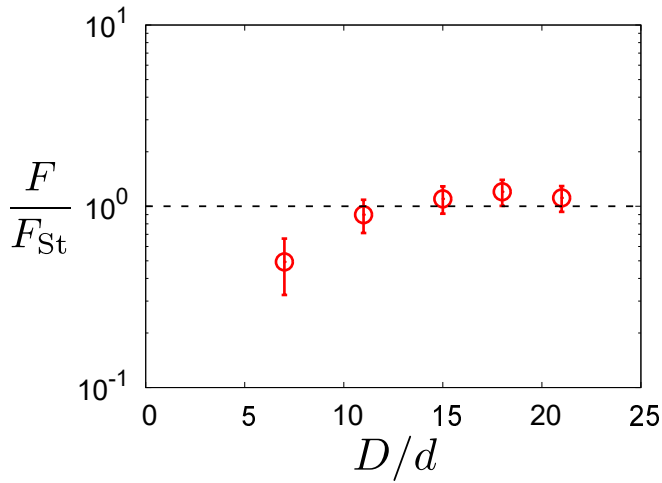}
	\caption{Plot of the drag force against the intruder diameter for $\phi=0.40$ and $V/v_{\rm T}=3.2\times 10^{-2}$.
	Here, $F_{\rm St}$ represents Stokes' law expressed in Eq.\ \eqref{eq:Stokes}.}
	\label{fig:periodic}
\end{figure}

\section{Diameter of the intruder dependence of the drag force for the periodic systems}\label{sec:periodic}
In the main text, we show how the drag force depends on the diameter when the beam particles are far from the intruder.
In this Appendix, however, we investigate the diameter dependence of the intruder on the drag force in the low-speed regime when the intruder is initially located inside stationary beam particles.
In addition, we adopt the periodic boundary condition in the flow direction to keep the packing fraction throughout the simulations in this Appendix.
The used parameters in this Appendix are listed in Table \ref{fig:parameters2}.

Figure \ref{fig:periodic} shows the drag force to Stokes' drag.
As the system size increases, the drag force approaches Stokes' law \eqref{eq:Stokes}.
This supports that the drag for the fully confined and periodic system is described by Stokes' law, which is consistent with the previous studies \cite{Vergeles95}.
The results in this Appendix are suggestive. 
Indeed, if we adopt the periodic boundary condition, Stokes' law is obtained without any difficulty. 
In other words, Epstein's law reported in the main text only appears as a transient from the initial hit of beam particles to approaching a steady flow.


\end{document}